# The Effects of Visual and Control Latency on Piloting a Quadcopter using a Head-Mounted Display

Jingbo Zhao  Robert S. Allison
Department of Electrical Engineering and Computer Science
York University
Toronto, Canada
{jingbo, allison}@cse.yorku.ca

Margarita Vinnikov
Department of Informatics
New Jersey Institute of Technology
Newark, USA
rita.vinni@gmail.com

Sion Jennings
Flight Research Laboratory
National Research Council
Ottawa, Canada
sion.jennings@nrc-cnrc.gc.ca

*Abstract*—Recent research has proposed teleoperation of robotic and aerial vehicles using head motion tracked by a head-mounted display (HMD). First-person views of the vehicles are usually captured by onboard cameras and presented to users through the display panels of HMDs. This provides users with a direct, immersive and intuitive interface for viewing and control. However, a typically overlooked factor in such designs is the latency introduced by the vehicle dynamics. As head motion is coupled with visual updates in such applications, visual and control latency always exists between the issue of control commands by head movements and the visual feedback received at the completion of the attitude adjustment. This causes a discrepancy between the intended motion, the vestibular cue and the visual cue and may potentially result in simulator sickness. No research has been conducted on how various levels of visual and control latency introduced by dynamics in robots or aerial vehicles affect users' performance and the degree of simulator sickness elicited. Thus, it is uncertain how much performance is degraded by latency and whether such designs are comfortable from the perspective of users. To address these issues, we studied a prototyped scenario of a head motion controlled quadcopter using an HMD. We present a virtual reality (VR) paradigm to systematically assess the effects of visual and control latency in simulated drone control scenarios.

## I. Introduction

Recently, there has been growing interest in remotely operating robots and aerial vehicles using head motion tracked by an HMD. A typical teleoperation approach for such vehicles maps the tracked head orientation by an HMD to the attitude of the vehicles for maneuvering it and first-person views from the perspective of the vehicles are usually captured by onboard cameras and presented onto the display panels of HMDs [1-4]. Such settings give users an egocentric, immersive and intuitive way of teleoperation compared with conventional control methods using hand-held devices, such as a joystick. However, a major difference between head motion control methods and conventional control methods is that, in the former case, head motion is coupled with visual updates. As the motion of a vehicle is constrained by its dynamics, appreciable visual and control latency always exists between the issue of control commands by head movements and the visual feedback received at the completion of the attitude adjustment. This causes a discrepancy between the intended motion, the vestibular cue and the visual cue and may potentially result in simulator sickness. In most VR applications, the major source of latency in HMDs comes from the motion-to-photon (end-to-end) latency [5]. However, motion-to-photon latency in HMDs, such as the Oculus Rift DK2, is estimated to be only a few milliseconds [6][7]. This is minimal compared to the latency introduced by dynamics in vehicles, which is dependent on the hardware and the controller used. Additional latency may be introduced if the head motion of the user is not properly mapped to the motion of vehicles. An example is the threshold technique, in which the vehicle starts to move or stop when Euler angles of a user's head pass certain thresholds, or the movement of the vehicle is triggered by certain gestures. We consider this as a type of discrete ON/OFF control input. A more responsive approach is to continuously map head motion to the motion of the vehicle so that the latency introduced in the motion mapping is minimal. For example, to fly a quadcopter, head orientation can be mapped to the tilt of the quadcopter and the quadcopter moves when it is tilted. We refer this type of operation as continuous inputs. Other sources of latency in head motion controlled vehicles include network transmission [8] and computations, but these were not considered in the present study as we focused on visual and control latency, which dominates in the envisioned scenarios. There has been no previous research on how various levels of visual and control latency introduced by dynamics in head motion controlled vehicles affect a user's performance and well-being. A common upper bound for tolerable latency in VR systems is usually taken as 20 ms [9] but visual and control latency introduced by dynamics is usually much higher. Thus, it is uncertain whether such head motion controlled vehicles affect performance and whether they are comfortable for users as these techniques may elicit simulator sickness. Therefore, in the present study, we simulated head motion controlled quadcopter scenarios using HMDs in a virtual environment. The VR paradigm enabled us to experimentally control the degree of latency and systematically assess the effects of latency in head motion controlled quadcopters.

The VR paradigm offers several advantages compared to experiments using real quadcopters. First, latency can be easily controlled by setting appropriate gains in the simulation equations of quadcopters. Second, we avoid losing or damaging a drone when users are unable to control the quadcopter due to technical failures or potential simulator sickness caused by experiments. Third, complex testing environments can be easily set up and flight data can be conveniently and accurately logged.

The goal of the present research is to assess users' flight performance and the degree of simulator sickness given various levels of visual and control latency introduced by the aerial dynamics of simulated quadcopters. We also aimed to investigate whether people feel comfortable teleoperating a quadcopter using an HMD and whether they can adapt to the latency with practice. The results of the research may serve as guidelines on the design of head motion controlled vehicles.

The rest of the paper is organized as follows: Section II discusses related work; Section III describes the experiment method; Section IV presents the experiment and the results; and Section V presents discussion based on the experiment results and draws the conclusion.

## II. RELATED WORK

### A. Joint Use of HMDs and Hand-held Devices for Controlling Vehicles

In these systems, head motions tracked by HMDs are usually only used for controlling camera orientations and the display panels of HMDs are used for viewing. Vehicles are controlled by hand-held devices, such as a joystick.

de Vries and Padmos [10] compared different viewing conditions for operating a camera on a simulated unmanned aerial vehicle (UAV) using head motions tracked by an HMD while piloting the UAV with a joystick. These conditions included:

- real HMDs *versus* simulated HMDs.
- stereoscopic viewing *versus* monoscopic viewing.
- fields of view.
- image lag, *etc*.

Morphew *et al*. [11] studied the performance of two methods for controlling a UAV:

- a joystick with an HMD.
- a joystick with a computer monitor.

Their study found that using a joystick with a computer monitor gave better task performance than using a joystick with an HMD and elicited less simulator sickness.

Mollet and Chellali [12] proposed the idea of using head tracking functionality in HMDs for teleoperation of cameras on robots to improve the degree of immersion by presenting the view from the robot vantage point on an HMD. Their prototypes were based on wheeled-robots and head rotation was used to control the attitude of an onboard camera. But the evaluations of the proposed system were not given in the paper.

Doisy *et al*. [13] compared the performance of three methods for controlling robots:

- an Xbox 360 controller for controlling both robot movement and camera orientation.
- an Xbox 360 controller for controlling robot movement and an HMD for controlling camera orientation.
- hand gestures for controlling robot movement and an HMD for controlling camera orientation.

Similar to the results in [11], they found that using the Xbox 360 controller for controlling robot movements and camera orientation had the best task performance.

A more recent work by Smolyanskiy and Gonzalez-Franco [14] presented a design of a quadcopter with stereoscopic viewing using the Oculus Rift DK2. Onboard camera panning was controlled digitally (as opposed to mechanically) by head motion while the quadcopter was piloted by a hand-held controller. Real flight tests were carried out to study the simulator sickness with an end-to-end latency of $250 \pm 30$ ms and participants had minor simulator sickness. Another experiment was conducted to study simulator sickness in relation to gaming experience and visual acuity using pre-recorded flight videos.

### B. Head Motion Controlled Vehicles using HMDs.

Several projects have explored teleoperation of robots and quadcopters using an HMD:

Martins and Ventura [1] presented a head motion controlled search and rescue (SAR) robot using an HMD. The HMD in the design had a three Degree-of-Freedom (DOF) tracker, which could capture yaw, pitch and roll angles of a user's head. The SAR robot had two tracks and an articulated frontal body with a stereo camera mounted on it. The yaw angle of a user's head was mapped to the angular velocity of the rotation of the SAR robot and the pitch angle was directly mapped to the angle of the front body of the robot to ascend or lower it. Since the robot was unable to perform roll movements, the roll angle of a user's head was used to rotate the image presented on the HMD. The experimental results showed that the control that utilized the HMD gave better performance in terms of depth perception, detail perception and the execution of a SAR operation compared to a 2D interface on a computer monitor with joystick control.

Higuchi and Rekimoto [2] proposed a control mechanism for quadcopters called the Flying Head. In this method, translation motion accomplished by walking and tracked by an HMD was directly mapped to the spatial motion of a quadcopter i.e. the $x$-, $y$- and $z$-positions and the yaw angle were synchronized between the user and the quadcopter. Experimental results showed that the proposed control method outperformed the conventional control method using a joystick when tracking static or moving targets.

Pittman and LaViola [3] evaluated six different techniques for operating a quadcopter, including five techniques based on head motion and body motion using an HMD and a technique using the Wiimote. To evaluate all six techniques, participants were asked to fly a quadcopter through five archways placed in an open environment. However, their results showed the Wiimote technique led to the shortest task completion time. Users had lower simulator sickness scores using the Wiimote interface compared with other interfaces. We note that in first five techniques, the control mechanisms were discrete ON/OFF control inputs. This may introduce higher latency and increase discrepancy between the intended motion, the vestibular cues and the visual cues. Thus, it is not surprising that performance of participants using the five head motion control techniques was lower compared to the Wiimote interface and the degree of simulator sickness was higher.

Teixeira *et al*. [4] used an augmented reality (AR) device – the Google Glasses for controlling a quadcopter. In this work, the gestures of the Google Glasses were mapped to the motion of the quadcopter and the video stream captured by the onboard camera of the drone was presented to the right eye of users. An advantage of the design is that operators know both their egocentric positions and the position of the quadcopter while operating the quadcopter. Thus, operators may simultaneously use other devices while flying the quadcopter. But the design may also confuse operators since visual inputs came from two different sources, which may potentially cause simulator sickness. Some preliminary evaluations regarding the interaction between operators and drones were conducted. But no details were given.

*C. Effects of Latency on Controlling Other Types of Vehicles*

Researchers have investigated the effects of latency on controlling vehicles, including cars and helicopters:

Blissing *et al.* [15] investigated the effects of visual latency on driving performance through an experiment conducted on real cars. Participants wore see-through HMDs modified from an Oculus Rift DK2 while driving. End-to-end latency was injected by delaying captured real-word video frames. The task was to drive a car through a slalom course under three different latency levels. Results showed that participants could compensate for increased latency by adapting their driving behaviours. But the tests were short-duration so simulator sickness would not be expected and it was not assessed.

Jennings *et al.* [16] studied the relationship between latency and flight performance of helicopter pilots. Two experiments were conducted to evaluate visual latency and control latency on piloting helicopters. These included asking pilots to fly a helicopter through two designated courses while wearing HMDs and perform precision hovering in a flight simulator. Results showed that both visual latency and control latency degraded flight performance. The frequency and the intensity of simulator sickness increased as longer delays were set.

*D. Summary*

In general, operating a vehicle with an HMD and a joystick does not provide performance advantages compared with operating a vehicle with a computer monitor and a joystick. The former [10-14] usually elicits symptoms of simulator sickness. However, operating vehicles using only HMDs [1-4] seem to give people a more direct, immersive and intuitive way of control than joint use of computer monitors and joysticks. In addition, researchers have also investigated the effects of latency on driving cars [15] and piloting helicopters [16]. However, no research has been conducted to study the effects of various levels of latency on head motion controlled quadcopters or other types of vehicles. The latency in the head motion controlled quadcopters differs with that of the helicopters. In helicopter control, the visual updates are coupled with the hand motions of pilots. Latency in such cases can be divided into visual latency and control latency. In head motion controlled quadcopters, visual updates and quadcopter motions are coupled with head motions and thus visual latency and control latency are combined into a single factor. Since using head motion tracked by an HMD to teleoperate a quadcopter or other robotic platforms is a very promising interface, the effects of visual and control latency in such scenarios are worth direct investigation.

## III. SIMULATION OF A HEAD MOTION CONTROLLED QUADCOPTER

*A. Software*

Visual and control latency for head motion controlled quadcopters is defined as the time interval between when a desired tilt angle of the quadcopter is given by the pitch and roll of a user's head and when the tilt angle of the quadcopter reaches the desired tilt angle. To isolate how visual and control latency elicits simulator sickness and affects flight performance, we focused on the simulation of a quadcopter that only allowed tilt (pitch and roll) and translation while yaw and altitude control were disabled. The application for simulating a head motion controlled quadcopter was developed using Python 2.7 based on the helicopter transport method in the Worldviz Vizard 5.0.

The position of a quadcopter flying at a fixed height can be represented by $P = (x, z)$. The attitude can be represented by $\Theta = (\theta, \Phi)$, where $\theta$ and $\Phi$ are the pitch and the roll of the quadcopter. The control input to a quadcopter can be defined as $u = (\theta_D, \Phi_D)$, which are the desired pitch and the desired roll of the quadcopter given by the roll and the pitch of a user's head tracked by an HMD. To translate, a quadcopter first tilts itself. The tilt of the quadcopter at a fixed height can be simulated by the following set of equations:

$$\begin{cases} \dot{\theta} = k_\theta(\theta_D - \theta) \\ \dot{\Phi} = k_\Phi(\Phi_D - \Phi) \end{cases} \quad (1)$$

The angular rates of the pitch and the roll ($\dot{\theta}$ and $\dot{\Phi}$) are proportional to the differences between the setpoints (pitch setpoint $\theta_D$ and roll setpoint $\Phi_D$) and their actual instantaneous values (pitch $\theta$ and roll $\Phi$). The constants of proportionality are the respective gains $k_\theta$ and $k_\Phi$. These gains control how fast the actual pitch $\theta$ and roll $\Phi$ converge to their setpoints $\theta_D$ and $\Phi_D$.

The simple dynamics equations allowed us to easily set the latency for simulation by modifying the gains $k_\theta$ and $k_\Phi$. We measured the latency as the rise time of the step responses of the tilted angles ($\theta$ and $\Phi$) from 0° to 10° given the corresponding step inputs from 0° to 10°. In Table I (see next page), we present the five different latency levels used in the experiment and their corresponding gains. The latency values for the experiments were within the normal range of the step response of attitude control of a quadcopter. For example, in [17], the measured rise time of pitch and roll were approximately 0.2 s and 0.4 s, respectively, when 10° step inputs were given during indoor flight. The gains were determined based on a rate of 75 Hz for updating the equations. However, it should be noted that the latency with a given gain is dynamic. For example, if a user tilts head for 1°, the duration for the quadcopter to adjust its attitude to reach the setpoint is much smaller than if the user tilts head for 10°. Thus, the latency is also dependent on users' behavior to tilt their heads and the gains only set the upper bounds for the latency values. When a quadcopter is tilted, it starts to translate. The translation motion is simulated by:

$$\begin{cases} \ddot{x}_B = (\Phi/\Phi_{max})a - \dot{x}_B d \\ \ddot{z}_B = (\theta/\theta_{max})a - \dot{z}_B d \end{cases} \quad (2)$$

The first terms in the equations of $\ddot{x}_B$ and $\ddot{z}_B$ show that the translation accelerations of the quadcopter are proportional to the tilt angles of the quadcopter scaled by $a$ ($a = 10 \, m/s^2$), which is the maximum translational acceleration of the quadcopter. Drag terms $\dot{x}_B d$ and $\dot{z}_B d$ – proportional to the velocities ($\dot{x}_B$ and $\dot{z}_B$) scaled by a factor $d$ ($d = 0.05$) – were introduced to give inertia to the acceleration of the quadcopter. The factor $d$ is not related to visual and control latency as it only damps the translation accelerations ($\ddot{x}_B$ and $\ddot{z}_B$). The maximum tilt angles $\Phi_{max}$ and $\theta_{max}$ of the quadcopter are limited ($\Phi_{max} = 10°$ and $\theta_{max} = 10°$). During real-time simulation, the accelerations and the angular rates were integrated over time

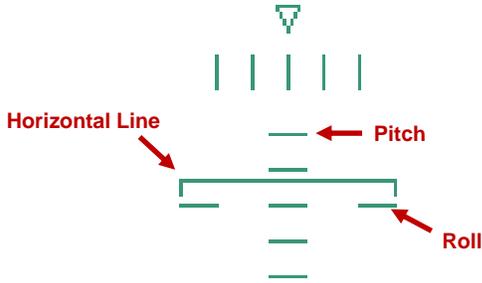

Fig. 1. HUD for the quadcopter

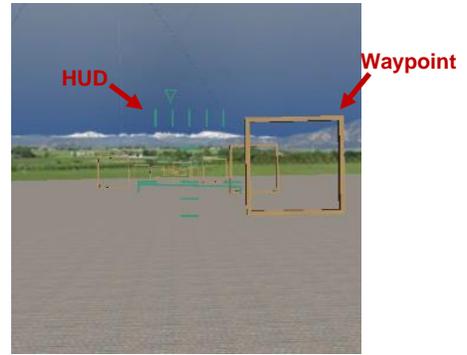

Fig. 2. Testing environment

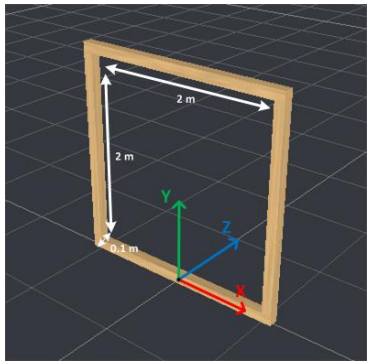

Fig. 3. Waypoint and its coordinate

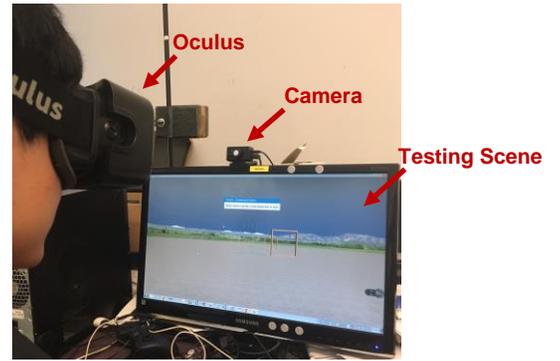

Fig. 4. Experiment setup

TABLE I. LATENCY LEVELS

| Latency Level | 1 | 2 | 3 | 4 | 5 |
|---|---|---|---|---|---|
| Gains $k_\theta$, $k_\Phi$ | 32.5 | 15.6 | 10.5 | 7.9 | 6.5 |
| Latency (s) | 0.2 | 0.4 | 0.6 | 0.8 | 1.0 |

TABLE II. EXPERIMENT SESSION ORDER

| Participant | Day 1 | Day 2 | Day 3 | Day 4 | Day 5 |
|---|---|---|---|---|---|
| P1 | Training, Latency 3 | Latency 5 | Latency 1 | Latency 2 | Latency 4 |
| P2 | Training, Latency 1 | Latency 4 | Latency 5 | Latency 3 | Latency 2 |
| P3 | Training, Latency 1 | Latency 2 | Latency 5 | Latency 3 | Latency 4 |
| P4 | Training, Latency 5 | Latency 3 | Latency 2 | Latency 1 | Latency 4 |
| P5 | Training, Latency 5 | Latency 2 | Latency 1 | Latency 3 | Latency 4 |
| P6 | Training, Latency 5 | Latency 4 | Latency 1 | Latency 3 | Latency 2 |
| P7 | Training, Latency 2 | Latency 3 | Latency 4 | Latency 1 | Latency 5 |
| P8 | Training, Latency 3 | Latency 4 | Latency 5 | Latency 2 | Latency 1 |
| P9 | Training, Latency 4 | Latency 2 | Latency 3 | Latency 1 | Latency 5 |

at the rate of 75 Hz and consequently the quadcopter moved in the virtual environment.

During our initial trials, we found that one difficulty operating the quadcopter from a first-person view was to judge the attitude of the quadcopter. People had difficulty in hovering the quadcopter and making precise maneuvers. To address this issue, we designed a simple Heads-Up Display (HUD), as shown in Fig. 1, which indicates the attitude of the quadcopter. The horizontal line represents the pitch by moving up and down and the roll by tilting.

We designed an experiment scene (Fig. 2) that contained a stone-textured ground and a skydome. Similar to the experiment environment in [3], which used archways, we placed 100 square waypoints (Fig. 3) in the scene at a height of 5 m and the altitude of the quadcopter was fixed at 6 m. The inner dimension of each waypoint was 2 m (W) × 2 m (H) × 0.1 m (D). The longitudinal distances (z-axis) between these waypoints were 5 m. The lateral positions (x-axis) were randomized in the interval of ± 5 m.

*B. Hardware*

We used the Oculus Rift DK2 to track the users' head motion and present the simulated drone control scenario. This HMD uses a hybrid optical-inertial tracker [18][19], which consists of an inertial measurement unit and a camera with an infrared lens, to track users' head motion at a sampling rate of approximately 75 Hz. The simulation was hosted on a computer running Windows 7 with an Intel i7 2.8 GHz CPU, 4 GB memory and an AMD Radeon HD 6850 graphics card. Fig. 4 presents the experiment setup.

## IV. EXPERIMENT

### A. Introduction

The goal of our experiment was to investigate the flight performance of the participants and the degree of simulator sickness as a function of various levels of latency. We hypothesized that higher latency would degrade flight performance and elicit more simulator sickness. We reasoned that flight performance may improve and simulator sickness may relieve across sessions. We also hypothesize that tolerance to the quadcopter scenario and flight performance may differ between participants.

We adopted a repeated-measures design for the experiment. The independent factor for experiment was the latency of dynamics controlled by setting the gains $k_\Theta$ and $k_\Phi$ in the simulation equations of the quadcopter in (1), which resulted in different latency values (Table I). To balance the order effect of treatments, we randomized the order of latency levels for each participant as shown in Table II.

### B. Participants

Ten people participated in the experiment. All had normal or corrected-to-normal vision and were naïve to the goals of the experiment. Informed consent was obtained from all participants in accordance with a protocol approved by the Human Participants Review Subcommittee at York University.

The experiments were conducted on five different days. On each experiment day, a participant completed one experiment session. The five experiment sessions for each participant were completed within two weeks. Simulator sickness questionnaires (SSQs) [20] were completed before and after each session.

### C. Procedure

During the experiment sessions, the participant wore the Oculus Rift DK2 and sat approximately 60 cm in front of the monitor, on which the camera of the HMD was attached. A researcher sat beside the participant and observed the virtual scene the participant saw on the computer monitor. At the beginning of an experiment session, a calibration procedure was conducted using the HUD (Fig. 1). Participants were asked to level their head by observing the horizontal line that indicated the pitch and the roll of their heads on the HUD such that the horizontal line overlapped with the three short horizontal lines. This indicated that the yaw and the pitch of the participant's head was close to zero. When calibration was completed, the researcher pressed the start button to initiate the flight and the data recording started immediately. The goal for participants was to pass through the centres of all waypoints and they were only allowed to move the quadcopter by pitch and roll movements. In case that they missed a waypoint, they were advised not to backtrack. Participants were required to enter a pink bounding volume (final destination) when they reached the last waypoint. Upon entering the pink bounding volume, the data recording stopped immediately, and the task was considered as completed. On the first day (as shown in Table II), a training session was conducted to teach participants the operation of the quadcopter using the HMD. The training scene was a simplified version of the actual experiment scene and only consisted of one waypoint and a pink bounding volume. The latency value was set to zero by directly mapping the pitch setpoint $\Theta_D$ and the roll setpoint $\Phi_D$ given by users' tracked head orientation to the pitch $\Theta$ and the roll $\Phi$ of the quadcopter, respectively. Pre-experiment SSQs were completed after the training session for the first day. The experiment application collected the timestamps, the positions and the Euler angles of the quadcopter for data analysis.

### D. Metrics

The data collected from the experiment enabled us to extract a wide range of parameters to assess the degree of simulator sickness and flight performance of participants. Specifically, we used the following parameters as the measures:

- **SSQ scores:**

The standard SSQ was used as the subjective measure to evaluate the degree of simulator sickness elicited by the experiment.

- **Task completion time $T$ ($s$):**

The duration from the start of the flight to the end of the flight.

- **Average flight speed $S$ ($m/s$):**

The average flight speed computed by dividing the total length of the flight path by task completion time $T$ ($s$).

- **Smoothness of the flight path $D$ ($m$):**

The mean value of the lateral distance (x-axis) of the actual flight path to the optimal flight path. We defined the optimal flight path as the shortest distance between the centres of two adjacent waypoints.

- **Number of waypoints passed $N_w$:**

The number of waypoints that participants successfully passed. Since we had 9 participants, the ideal value of the total number $S_w$ of the waypoints passed by all participants was nine hundred.

- **Number of collisions $N_c$:**

The number of the collisions with the frames of the waypoints. We assumed the quadcopter had a dimension of 0.3 m (W) × 0.1 m (H) × 0.3 m (D), which is a typical size of a civilian quadcopter. Participants were required to pass through the centers of waypoints as they were unable to judge whether the quadcopter would collide with the waypoints from the first-person view. The collisions were determined by computing the intersection of the bounding boxes of the quadcopter and the frames of the waypoints. The total number of collisions by all participants is denoted as $S_c$.

### E. Results

One person withdrew from the experiment after attending the training session due to simulator sickness. Thus, we had valid data from nine participants (age: 27.1 ± 5.5, 7 males, 2 females).

We grouped the extracted parameters by latency levels and by the order of experiment sessions to show whether the extracted parameters are more related to latency levels or session orders (error bars in Fig. 5 – Fig. 12 denote standard deviation).

Since the SSQ scores either remained the same or increased after the completion of an experiment session, we took absolute

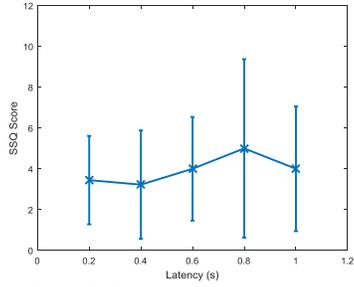

Fig. 5. SSQ score increase by latency

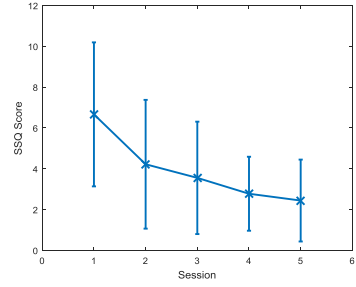

Fig. 6. SSQ score increase by session

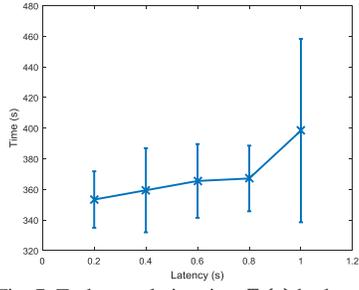

Fig. 7. Task completion time $T\ (s)$ by latency

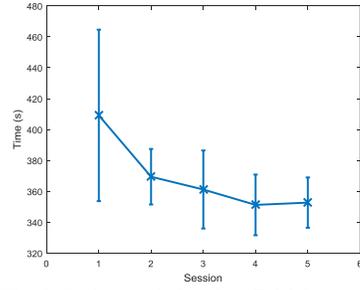

Fig. 8. Task completion time $T\ (s)$ by session

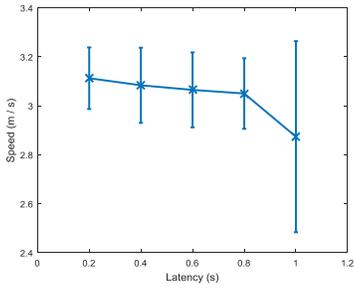

Fig. 9. Average flight speed $S\ (m/s)$ by latency

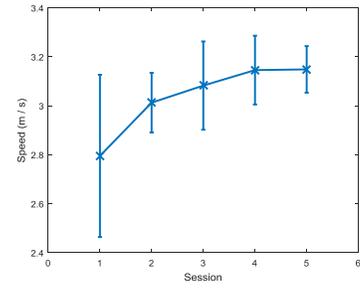

Fig. 10. Average flight speed $S\ (m/s)$ by session

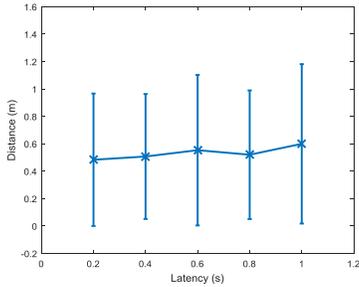

Fig. 11. Path smoothness $D\ (m)$ by latency

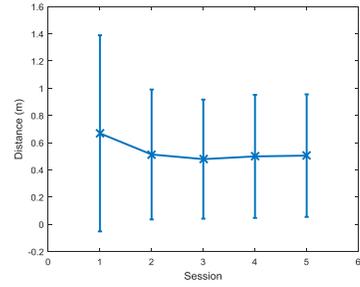

Fig. 12. Path smoothness $D\ (m)$ by session

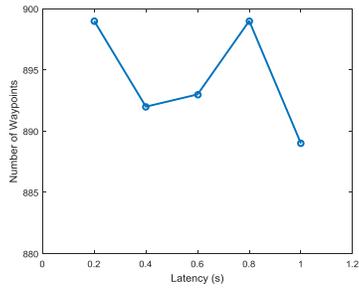

Fig. 13. Total number of waypoints passed $S_w$ by latency

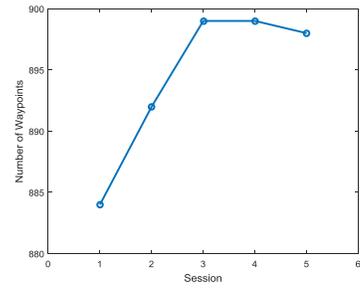

Fig. 14. Total number of waypoints passed $S_w$ by session

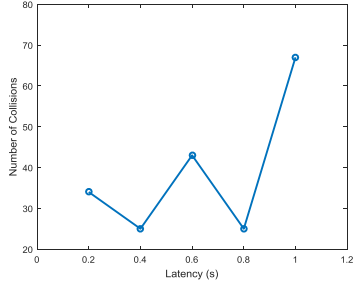

Fig. 15. Total number of collisions $S_c$ by latency

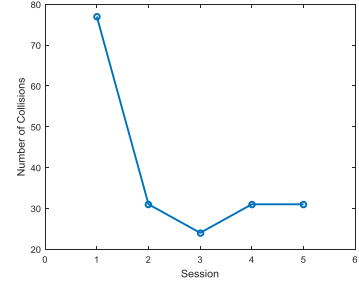

Fig. 16. Total number of collisions $S_c$ by session

TABLE III. RESULTS OF THREE-WAY ANOVA ANALYSES (SIGNIFICANT $P$-VALUES ARE SHADED)

|  |  | $SSQ$ | $T\ (s)$ | $S\ (m/s)$ | $D\ (m)$ | $N_w$ | $N_c$ |
|---|---|---|---|---|---|---|---|
| *Participant* | $F(8, 28)$ | 8.93 | 3.60 | 3.47 | 16.29 | 2.41 | 2.54 |
|  | $p$ | <0.001 | 0.005 | 0.007 | <0.001 | 0.04 | 0.03 |
| *Session* | $F(4, 28)$ | 8.85 | 6.96 | 6.33 | 6.19 | 1.25 | 1.7 |
|  | $p$ | <0.001 | <0.001 | <0.001 | 0.001 | 0.31 | 0.18 |
| *Latency* | $F(4, 28)$ | 2.09 | 3.12 | 2.04 | 1.56 | 0.49 | 0.97 |
|  | $p$ | 0.11 | 0.03 | 0.12 | 0.21 | 0.74 | 0.44 |

scores for all sessions to remove pre-experiment biases. In general, SSQ scores were high when latency levels were high (Fig. 5), which showed that high latency values increased simulator sickness. Because of the continuous mapping between the head orientation and the tilt of the quadcopter, the degree of simulator sickness was subtle for all five latency conditions. In terms of sessions, we found that people adapted across experiment sessions and the SSQ scores gradually declined as more experiment sessions were completed by participants (Fig. 6).

High latency values also increased task completion time $T\ (s)$ (Fig. 7) and reduced average flight speed $S\ (m/s)$ (Fig. 9). There were two reasons. First, when latency was high, the tilt motion of the quadcopter is slow. Hence, the acceleration of the quadcopter was also slow as the acceleration is dependent on the tilt of the quadcopter. Second, when latency was high, participants could have much difficulty in maneuvering the quadcopter. Participants may have to perform more rolls to point the quadcopter to the centre of a waypoint before flying it through. Due to learning and adaptation effects, task completion time gradually shortened (Fig. 8) and average flight speed also increased (Fig. 10) as participants attended more experiment sessions.

Different latency conditions slightly affected the smoothness of the flight path $D\ (m)$ (Fig. 11). High latency generally led to less smooth flight paths. We also found that as more experiment sessions were conducted, flight path also became smoother (Fig. 12). This indicated that participants performed fewer rolls, which also can be attributed to learning and adaptation effects.

The number of waypoints passed $N_w$ and the number of collisions $N_c$ were coarse parameters compared with previous four parameters. These two parameters were only affected when participants deviated too much from the centres of the waypoints, which resulted in missing or colliding with the waypoints. We did not observe meaningful trends to plot the number of waypoints passed $N_w$ and the number of collisions $N_c$ in terms of mean value and standard deviation so the total number of waypoints passed $S_w$ and the total number of collisions $S_c$ were plotted instead (Fig. 13 – Fig. 16). Although we did not find clear patterns when grouping these two parameters by latency (Fig. 13 and Fig. 15), both parameters improved in terms of sessions (Fig. 14 and Fig. 16), which indicated that the participants as a group became more skilled in controlling quadcopters with more practice.

In general, when latency was higher, the standard deviation of an extracted parameter was usually large. This showed that the tolerance of high latency differed between participants. Similarly, we also observed that the standard deviations of the extracted parameters of the first sessions were high. This indicated that piloting skills varied between participants. Some people were initially good at piloting a quadcopter using an HMD while others were not. Standard deviations decreased in later sessions, which indicated the performance between participants became similar.

In Table III, we summarized the results of the three-way ANOVA analyses using the function anovan() in Matlab to study the effects of the extracted parameters. The model included *Session* and *Latency* as fixed factors while *Participant* was treated as a random factor. This is similar to the model presented in [14]. A detailed explanation of the mixed model ANOVA that included fixed factors and random factors can be found in [21]. In general, *Participant* had significant effects on all parameters, which demonstrated that tolerance and skills to the quadcopter control scenario differed between participants. *Session* had significant effects on SSQ scores, task completion time $T\ (s)$, average flight speed $S\ (m/s)$ and smoothness of the flight path $D\ (m)$ due to learning and adaptation. *Latency* had a significant effect on task completion time $T\ (s)$. One reason that *Latency* did not influence other parameters was that participants consistently adapted to the quadcopter scenario as they attended more experiment sessions and the latency effects were therefore mitigated. Second, visual and control latency is dynamic and dependent on how users tilt their heads. Moving heads in small angles with slow motion results in lower latency compared to making abrupt head movements in large angles. We observed that participants adopted the strategy to move their heads slower and in smaller angles as they attended more sessions.

Pairwise comparisons on fixed factors were conducted using Tukey's range tests. These tests showed that for the factor *Session*, there were significant differences between the first session and the rest four sessions on SSQ scores, task completion time $T\ (s)$ and smoothness of the flight path $D\ (m)$. A significant difference was also found on average flight speed $S\ (m/s)$ between the first session and the last three sessions. These results may suggest that participants made substantial progress in adapting to the quadcopter scenario after they completed the first session. In terms of the factor *Latency,* there was a significant difference between the first session and the fifth session on task completion time $T\ (s)$, which showed that high latency increased task completion time $T\ (s)$.

V. DISCUSSION AND CONCLUSION

In this paper, we presented a VR experiment to systematically evaluate the effects of visual and control latency introduced by dynamics of a simulated quadcopter. We showed that a latency value with an upper bound of 1 s only elicited subtle simulator sickness when head motion was continuously mapped to the motion of the quadcopter. In addition, high latency values resulted in worse flight performance and higher level of simulator sickness. We also showed that as participants attended more experiment sessions, they became more tolerant to the head motion controlled drone scenarios. Lower SSQ scores were reported and flight performance also improved. These results have verified our hypotheses that higher latency would degrade flight performance and elicit more simulator sickness and that tolerance to the quadcopter scenario and flight performance differ between participants.

The present study suggests piloting a quadcopter using an HMD is feasible in terms of tolerance to visual and control latency, but training is needed for people to efficiently and comfortably operate the quadcopter with the interface. Selecting people inherently good at the quadcopter control scenario as pilots facilitates the training process. In addition, using a quadcopter that has a fast response to changing tilt commands would both improve the flight performance and reduce simulator sickness. Additional recommendations in designing such systems are to use the HMDs with low motion-to-photon latency and to continuously map head motion to quadcopter motion.

In [22], researchers presented a user study that assessed simulator sickness using videos presented on an HMD that simulated a hovering quadcopter disturbed by winds. Further research should study flight performance and simulator sickness when a quadcopter is subject to air disturbance when users perform flight maneuvers using VR or on flights of real quadcopters. In addition, the influence of the yaw and the altitude control of a quadcopter on flight performance and simulator sickness also needs to be studied. In the current study, the scene setup was relatively simple, so it also will be interesting to investigate the relationship between scene complexity and simulator sickness. Finally, an experiment to study visual and control latency with real quadcopters also needs to be conducted.